\documentclass[aps,preprint]{revtex4}%
\usepackage{amsfonts}
\usepackage{amsmath}
\usepackage{amssymb}
\usepackage{graphicx}%
\begin{document}
\preprint{cond-mat}
\title[Short title for running header]{Phase diffusion in stationary state of nonequilibrium Bose gas. }
\author{V.S. Babichenko}
\affiliation{RSC "Kurchatov Institute", Moscow 123182, Kurchatov Square 1, Russia}
\pacs{PACS number}

\begin{abstract}
The low energy properties of the stationary state of a nonequilibrium Bose gas
with the dynamical equilibrium between the escape and creation of particles
are studied. The low energy spectrum of elementary excitations in such Bose
gas is calculated at sufficiently low temperatures. Both in the nonequilibrium
stationary state and in the thermodynamically equilibrium state the spectrum
of the low energy excitations is formed by the long wavelength phase
fluctuations of the complex Bose field. The spectrum of the phase fluctuations
is found to have a diffusion behavior. Such behavior is in contrast to that in
the thermodynamically equilibrium BEC systems in which the long wavelength
phase fluctuations have the sound spectrum. The diffusion character of the
phase fluctuations is due to noise generated by the escape and creation of
particles. The diffusion character of the phase fluctuation spectrum results
in the absence of coherency and, thus, in the absence of BEC and superfluidity
as well.

\end{abstract}
\volumeyear{ }
\volumenumber{ }
\issuenumber{ }
\eid{ }
\date{}
\received[Received text]{}

\revised[Revised text]{}

\accepted[Accepted text]{}

\published[Published text]{}

\startpage{0}
\endpage{1}
\maketitle

\section{Introduction.}

The investigations of the coherent properties of a Bose gas of quasiparticles
in the nonequilibrium systems attract large interest of experimentalists and
theorists. The various efforts to realize the state with the Bose-Einstein
condensate (BEC) in the nonequilibrium Bose gas of quasiparticles were carried
out in different systems. The systems are, as follows, exciton gas in the
double quantum wells in semiconductors \cite{BEC}-\cite{Lev}, polariton
systems, in particular, polariton systems in the two dimensional (2D) cavities
located between two mirrors \cite{MCav}-\cite{KeelBerl}, and a gas of excited
magnons \cite{Dem}-\cite{Vol}. All these systems are the candidates to observe
BEC at relatively large temperatures. Some of them are interesting both for
the fundamental and for the applied science. In all these systems
quasiparticles are created by external forces and live during a finite time.
In what follows, quasiparticles are called as particles for brevity.

The properties of the stationary state of a nonequilibrium Bose gas produced
by the dynamic equilibrium between the escape and creation of particles due to
external forces are considered in the present paper. In contrast to quantum
atomic gases in traps in which BEC has been observed at ultralow temperatures
and where the atoms have the infinite lifetime \cite{BEC}, \cite{Pit}, all the
systems considered here have a finite lifetime for Bose particles. Provided
the lifetime is long compared with the thermalization time, the system can be
observed within the time interval small compared with the particle lifetime
and large compared with the thermalization time. In this case the external
pumping is unnecessary. However, if the lifetime is small compared with the
thermalization time, a continuous external pumping of particles becomes
necessary to reach a stationary state of the system. Below we analyze this
case within the framework of the nonideal Bose gas model augmented with the
processes of the continuous escape and creation of particles. Moreover, the
latter processes are supposed to be noncoherent in accordance with many
experiments, for example, \cite{B1}-\cite{Tim2}, \cite{Kaspr1}-\cite{KenFor}.
In particular, for polaritons in microcavities, a pumping produces particles
with the large energies and then the particles relax to the low energy states.
We assume that the processes of the escape and creation of particles are
governed by random sources. In the stationary state the escape and creation
terms balance each other in average but the noise generated by these two
processes acts on the system continuously.

It is well known that the possibility of BEC and superfluidity is closely
connected with the behavior of the low energy part of the excitation spectrum.
The low energy spectrum of elementary excitations is found here for the
stationary state of a nonequilibrium Bose gas at sufficiently low
temperatures. In the nonequilibrium Bose gas in the stationary state as well
as in the thermodynamically equilibrium gas with BEC the renormalized low
energy spectrum of elementary excitations is governed by the long wave phase
fluctuations of the complex Bose field. As is well known, the low energy
spectrum of phase fluctuations and, correspondingly, the low energy part of
the excitation spectrum in the thermodynamically equilibrium Bose gas with BEC
has a sound-like behaviour \cite{Bogol}-\cite{AGD}. We show here that, in
contradiction with the thermodynamically equilibrium Bose systems, the
spectrum of long wave phase fluctuations for the stationary state in the
nonequilibrium Bose gas with the noncoherent particle creation and escape has
a diffusion character at sufficiently small temperatures. The diffusion
behaviour of the spectrum is due to the noise generated by the escape and
creation processes. This noise exists in spite of the balance in average
between the escape and creation terms. The frequency behavior for the
different components of the self-energy part of the one-particle Green
function \cite{Schw}, \cite{Keld}, \cite{LLK} determines the influence of the
noise on the system. The relation between the kinetic and imaginary parts of
the retarded or advanced components in the self-energy part gives an effective
temperature of the stationary state. The important difference between
thermodynamically equilibrium Bose systems and nonequilibrium systems in the
stationary state results from nonzero value of the zero-frequency Fourier
component in the kinetic term of the one-particle self-energy part of
nonequilibrium systems. This circumstance does lead to the diffusion character
of the low energy spectrum of the phase fluctuations in the nonequilibrium
stationary state of a Bose gas. At the same time, the fluctuations of the Bose
field module has the same character as in the thermodynamically equilibrium
state of Bose gas. The diffusion character of the spectrum of phase
fluctuations results in the absence of coherency and, thus, in the absence of
BEC and superfluidity. In particular, for the systems with phase diffusion,
the interference experiments should not demonstrate the pronounced
interference picture in contrast to similar experiments in the quantum atomic
Bose gases in traps \cite{AnKett}, \cite{Pit}. Unlike thermodynamically
equilibrium Bose gas with BEC, the classical Bose field for the stationary
state of nonequilibrium Bose gas obeys the Fokker-Planck equation which is
more complicated than the Gross-Pitaevskii one. This is completely due to an
existence of the noise generated by the particle escape and creation. So, if
we try to describe the system by the Gross-Pitaevskii equation, the noise term
should be augmented to the right-hand side of the Gross-Pitaevskii equation.
As a result, the equation goes over to the Langevin-like equation.

At present, a majority of the systems used for the experimental
BEC\ realization in nonequilibrium stationary state of quasiparticles have the
quasi 2D geometrical configurations. For example, polaritons in microcavities
are confined by two parallel mirror planes with a small spacing between them.
In this case the bare spectrum of polaritons has a quadratic dispersion law in
the region of sufficiently small momenta near the bottom of the spectrum. For
the sufficiently small densities, such polariton systems can be described by
the nonlinear Schroedinger Hamiltonian with a noncoherent pumping and escape
of particles \cite{MCav}, \cite{MCav1}. The experiments on the BEC realization
or laser effect in the polariton systems in microcavities are carried out
actively at the present time \cite{Kaspr1}-\cite{KenFor}.

In the present paper the coherent properties and the low energy spectrum of
excitations in the stationary state of nonequilibrium Bose gas are studied in
the assumption that the system considered can be described by the nonlinear
Schroedinger action with a noncoherent pumping and escape of particles. As
will be seen below, in such systems the renormalized low energy spectrum has a
diffusion character both in two and in three dimensions. To be closer to
experiments \cite{Kaspr1}-\cite{KenFor} in which the BEC of quasiparticles has
been investigated, we consider the properties of a nonequilibrium Bose gas in
the 2D geometrical configuration with a brief analysis for the typical
properties of this dimensionality.

\section{Low energy spectrum of a nonequilibrium Bose gas in the stationary
state.}

In the present paper we study the coherent properties of the stationary state
in dilute 2D polariton gas in a microcavitiy under continuous noncoherent
pumping and escape of particles at small effective temperatures. These
properties are determined by the particles with the momenta near the bottom of
the polariton spectrum. The polaritons are concentrated in the region near the
bottom of the spectrum after relaxation from the excited states of particles
created by the noncoherent pumping. In the region of sufficiently small
momenta the bare polariton spectrum has a quadratic dependence on the momentum
for polariton gas in a microcavity confined by two plane mirrors separated by
sufficiently small spacing \cite{MCav}, \cite{MCav1}. The renormalized
spectrum of elementary excitations in the region of small momenta is
determined by an effective interparticle interaction. In this region of
momenta the polariton system of sufficiently small density can be considered
as a dilute 2D Bose gas with the effective interparticle interaction
independent of the momentum transfer. The Hamiltonian of such Bose gas with
the omitted terms of the escape and pumping of particles can be taken in the
form of the nonlinear Schroedinger Hamiltonian%

\begin{equation}
\widehat{H}=\int d^{2}r\left\{  \widehat{\psi}^{+}\left(  -\frac{1}%
{2m}\overrightarrow{\nabla}^{2}-\mu\right)  \widehat{\psi}+\frac{1}{2}g\left(
\widehat{\psi}^{+}\widehat{\psi}\right)  ^{2}\right\}  \label{H}%
\end{equation}

where $g>0$ is the two-particle scattering amplitude independent of the
momentum transfer, $\widehat{\psi}^{+}$, $\widehat{\psi}$ are the creation and
annihilation operators of Bose particles, $m$ is the effective mass, and $\mu$
is the chemical potential. We put here $\hbar=1$. The particle density
$n=<\widehat{\psi}^{+}\widehat{\psi}>$ is supposed to be small. In the case of
the low density Bose gas the chemical potential is $\mu=gn$. The consideration
of low energy excitations gives the possibility to take the bare spectrum of
particles in the form $\overrightarrow{p}^{2}/2m$, or in the coordinate
representation it can be written as $\left(  -\overrightarrow{\nabla}%
^{2}/2m\right)  $. The effect of the escape and creation processes will be
included into the description of the system via introducing the relaxation
terms into the Gross-Pitaevskii action treated within the framework of the
Keldysh-Schwinger technique \cite{Schw}, \cite{Keld}, \cite{LLK} in the
functional-integral formulation \cite{BK}.

In the "tetragonal" representation of the Keldysh-Schwinger technique the
action of the system $S$ can be written in the terms of integrating over the
double time contour with time reversion. The particle escape and creation
processes can be included into action $S$ as random sources of the particle
escape and creation. The action can be written as%

\begin{equation}
S=%
%TCIMACRO{\doint }%
%BeginExpansion
{\displaystyle\oint}
%EndExpansion
dtd^{2}r\left\{
\begin{array}
[c]{c}%
\overline{\psi}\left(  i\partial_{t}+\mu-\frac{1}{2m}\overrightarrow{\nabla
}^{2}\right)  \psi-\frac{1}{2}g\left(  \overline{\psi}\psi\right)  ^{2}-\\
-%
%TCIMACRO{\dsum \limits_{\alpha=1,2}}%
%BeginExpansion
{\displaystyle\sum\limits_{\alpha=1,2}}
%EndExpansion
\left(  \left(  \overline{\psi}f_{\alpha}+\overline{f}_{\alpha}\psi\right)
+\overline{f}_{\alpha}\widehat{K}_{\alpha}f_{\alpha}\right)
\end{array}
\right\}  \label{S4}%
\end{equation}

where $\psi$ and complex conjugate $\overline{\psi}$ are the Bose fields,
$f_{\alpha}$, $\overline{f}_{\alpha}$ are random sources depending on the time
and space coordinates, index $\alpha=1$ corresponds to the particle escape and
$\alpha=2$ corresponds to the particle creation by the external forces, and
$\widehat{K}_{\alpha}$ is the white noise correlator of random sources
$f_{\alpha}$. The corresponding generating functional Z with the action S
(\ref{S4}) is written in the form of the functional integral over the fields
$\psi$ and $f_{\alpha}$%

\begin{equation}
Z=\int D\psi D\overline{\psi}%
%TCIMACRO{\dprod \limits_{\alpha=1,2}}%
%BeginExpansion
{\displaystyle\prod\limits_{\alpha=1,2}}
%EndExpansion
Df_{\alpha}D\overline{f}_{\alpha}\exp\left\{  iS+i\delta S\right\}  \label{Z}%
\end{equation}

where the action $\delta S$ depends on the infinitesimal sources $\zeta$ as%

\begin{equation}
\delta S=%
%TCIMACRO{\doint }%
%BeginExpansion
{\displaystyle\oint}
%EndExpansion
dtd^{2}r\left(  \overline{\psi}\zeta+\overline{\zeta}\psi\right)  \label{delS}%
\end{equation}

The transition to the "triangular" representation can be performed by the substitution%

\begin{align}
\Theta &  =\frac{1}{2}\left(  \psi_{+}+\psi_{-}\right) \label{tr}\\
\theta &  =\psi_{+}-\psi_{-}\nonumber
\end{align}

where $\psi_{+}$ and $\psi_{-}$ are the Bose fields at the upper and lower
branches of the time contour. The component $\Theta$\ is the "classical"
component of the field and $\theta$ is the component corresponding to quantum
fluctuations. Note that the component $\Theta$ is nonzero for the case of the
fields coinciding at the different branches of the time countour $\psi
_{+}\left(  t,\overrightarrow{r}\right)  =\psi_{-}\left(  t,\overrightarrow
{r}\right)  $ , and the component $\theta$\ vanishes in this case. The
generating functional for low energy particles can be treated in the quasi
classical approximation due to the large value of the low energy state
occupation numbers. In this approximation the terms $\overline{\theta}\theta$
can be neglected as compared with the terms $\overline{\Theta}\Theta$. The
transition to the fields $\Theta$, $\theta$\ and the integration over the
random sources $f_{\alpha}$ in the generating functional (\ref{Z})\ with
taking the inequality $\overline{\theta}\theta<<\overline{\Theta}\Theta$ into
account give the generating functional for the low energy fields. The
corresponding action $S_{QCl}\left[  \Theta,\theta\right]  $ in the
"triangular" representation \cite{Schw}, \cite{Keld} reads%

\begin{equation}
Z=%
%TCIMACRO{\dint }%
%BeginExpansion
{\displaystyle\int}
%EndExpansion
D\Theta D\overline{\Theta}D\theta D\overline{\theta}\exp\left\{
iS_{QCl}\left[  \theta,\Theta\right]  +i\delta S_{QCl}\left[  J,j\right]
\right\}  \label{ZQCl0}%
\end{equation}

\begin{equation}
S_{QCl}\left[  \theta,\Theta\right]  =%
%TCIMACRO{\dint }%
%BeginExpansion
{\displaystyle\int}
%EndExpansion
dtd^{2}r\left\{
\begin{array}
[c]{c}%
\overline{\theta}\left(  i\partial_{t}+\mu-\frac{\widehat{\overrightarrow{p}%
}^{2}}{2m}+i\widehat{\Gamma}_{R}\left(  \widehat{\omega}\right)  \right)
\Theta+\\
+\overline{\Theta}\left(  i\partial_{t}+\mu-\frac{\widehat{\overrightarrow{p}%
}^{2}}{2m}+i\widehat{\Gamma}_{A}\left(  \widehat{\omega}\right)  \right)
\theta-\\
-g\overline{\Theta}\Theta\left(  \overline{\Theta}\theta+\overline{\theta
}\Theta\right)  +i\overline{\theta}\Gamma_{K}\left(  \widehat{\omega}\right)
\theta
\end{array}
\right\}  \label{SQCl0}%
\end{equation}

The frequencies and gradients of the fields describing low energy excitations
and producing the action $S_{QCl}\left[  \theta,\Theta\right]  $ are supposed
to be small on the scale of $\mu=g<\overline{\Theta}\Theta>=gn$, where $\mu$
is the chemical potential, $\widehat{\overrightarrow{p}}=-i\overrightarrow
{\nabla}$, $\widehat{\omega}=i\partial_{t}$ are the momentum and frequency
operators in the space-time representation. The frequency-dependent terms
$\widehat{\Gamma}_{R}\left(  \omega\right)  $, $\widehat{\Gamma}_{A}\left(
\omega\right)  $, $\widehat{\Gamma}_{K}\left(  \omega\right)  $ are the
imaginary parts of the corresponding self-energy part components of
one-particle Green function. The indices R, A, K denote the retarded, advanced
and kinetic components, respectively. These terms can be expressed via the
relaxation parts corresponding to the escape $\widehat{\Gamma}^{\left(
out\right)  }\left(  \omega\right)  $ and creation $\widehat{\Gamma}^{\left(
in\right)  }\left(  \omega\right)  $ of particles%

\begin{align}
\widehat{\Gamma}_{R}\left(  \omega\right)   &  =\widehat{\Gamma}_{R}^{\left(
out\right)  }\left(  \omega\right)  -\widehat{\Gamma}_{R}^{\left(  in\right)
}\left(  \omega\right) \label{GamInOut}\\
\widehat{\Gamma}_{A}\left(  \omega\right)   &  =\widehat{\Gamma}_{A}^{\left(
out\right)  }\left(  \omega\right)  -\widehat{\Gamma}_{A}^{\left(  in\right)
}\left(  \omega\right) \nonumber\\
\widehat{\Gamma}_{K}\left(  \omega\right)   &  =\widehat{\Gamma}_{K}^{\left(
out\right)  }\left(  \omega\right)  +\widehat{\Gamma}_{K}^{\left(  in\right)
}\left(  \omega\right) \nonumber
\end{align}

For the stationary state, the relaxations $\widehat{\Gamma}_{R}\left(
\omega\right)  $, $\widehat{\Gamma}_{A}\left(  \omega\right)  $ vanish at zero
frequency $\widehat{\Gamma}_{R}\left(  0\right)  =\widehat{\Gamma}_{A}\left(
0\right)  =0$. In contrast, the term $\widehat{\Gamma}_{K}\left(  0\right)
=\widehat{\Gamma}_{K}^{\left(  out\right)  }\left(  0\right)  +\widehat
{\Gamma}_{K}^{\left(  in\right)  }\left(  0\right)  $ is positive and nonzero
since this term is a sum of two positive nonzero terms $\widehat{\Gamma}%
_{K}^{\left(  out\right)  }\left(  0\right)  $ and $\widehat{\Gamma}%
_{K}^{\left(  in\right)  }\left(  0\right)  $ \cite{Schw}. The kinetic
component of the self-energy part $\widehat{\Gamma}_{K}$ describes the noise
generated by the particle escape and creation. For small frequencies $\omega$,
or in the time representation for small time derivative $\widehat{\omega
}=i\partial_{t}$, the relaxation terms can be expanded in $\omega$. The zero
frequency components $\widehat{\Gamma}_{R}\left(  0\right)  $ and
$\widehat{\Gamma}_{A}\left(  0\right)  $ have zero value in the stationary
state due to the dynamic balance of the particle escape and creation
processes. For this reason, in the limit of small frequencies the relaxations
$\widehat{\Gamma}_{R}\left(  \omega\right)  $, $\widehat{\Gamma}_{A}\left(
\omega\right)  $ can be expanded into a series in $\omega$. The first term of
the expansion has the form%

\begin{equation}
\widehat{\Gamma}_{R}\left(  \omega\right)  =-\widehat{\Gamma}_{A}\left(
\omega\right)  =\varkappa\omega=\varkappa i\partial_{t} \label{GRA}%
\end{equation}

where the constant $\varkappa$ can be written as a ratio of $\Gamma_{K}\left(
0\right)  $ to some constant $T^{\ast}$, which can be interpreted as an
effective temperature of the stationary state of the system%

\begin{equation}
\varkappa=\frac{\Gamma_{K}}{4T^{\ast}} \label{Kap}%
\end{equation}

For small $\omega$, the dependence of $\Gamma_{K}\left(  \omega\right)  $\ on
frequency $\omega$\ can be neglected as the zero frequency component of this
term is nonzero $\Gamma_{K}\left(  0\right)  =\Gamma_{K}\neq0$.\ The effective
temperature $T^{\ast}$ is supposed to be much smaller than the
Berezinskii-Kosterlitz-Thouless temperature $T_{c}$ \cite{Berez}, \cite{KT},
\cite{Popov}. Moreover, we suppose that $T^{\ast}<<$ $\mu$.\ The infinitesimal
term $\delta S_{QCl}^{\left(  pol\right)  }\left[  J,j\right]  $ is
proportional to the infinitesimal sources $J$ and $j$ obtained from the
infinitesimal sources $\zeta$ by the transformation (\ref{tr}). In this case
the sources $\zeta_{+}$, $\zeta_{-}$ play a role of the fields $\psi_{+}$,
$\psi_{-}$%

\begin{equation}
\delta S_{QCl}\left[  J,j\right]  =%
%TCIMACRO{\dint }%
%BeginExpansion
{\displaystyle\int}
%EndExpansion
dtd^{2}r\left[  \left(  \overline{\theta}J+\overline{J}\theta\right)  +\left(
\overline{\Theta}j+\overline{j}\Theta\right)  \right]  \label{DeltaS}%
\end{equation}

Using notations (\ref{GRA}) and (\ref{Kap}), we can rewrite\ the action
$S_{QCl}\left[  \theta,\Theta\right]  $ as%

\begin{equation}
S_{QCl}\left[  \theta,\Theta\right]  =%
%TCIMACRO{\dint }%
%BeginExpansion
{\displaystyle\int}
%EndExpansion
dtd^{2}r\left\{
\begin{array}
[c]{c}%
\overline{\theta}\left[  i\left(  1+i\varkappa\right)  \partial_{t}%
\Theta-F\right]  +\\
+\left[  -i\left(  1-i\varkappa\right)  \partial_{t}\overline{\Theta
}-\overline{F}\right]  \theta+i\overline{\theta}\Gamma_{K}\theta
\end{array}
\right\}  \label{SQCl1}%
\end{equation}

where the functions $F$ and $\overline{F}$ are%

\begin{align}
F  &  =\left(  \frac{\widehat{\overrightarrow{p}}^{2}}{2m}+g\overline{\Theta
}\Theta-\mu\right)  \Theta\label{F0}\\
\overline{F}  &  =\left(  \frac{\widehat{\overrightarrow{p}}^{2}}%
{2m}+g\overline{\Theta}\Theta-\mu\right)  \overline{\Theta}\nonumber
\end{align}

One can see that the functions $F$ and $\overline{F}$ can be represented in
the form of the functional derivatives%

\begin{equation}
F=\widehat{\overline{\partial}}H_{0}\text{; \ \ \ }\overline{F}=\widehat
{\partial}H_{0}\text{\ } \label{FH0}%
\end{equation}

where $\widehat{\partial}=\delta/\delta\Theta$ and $\widehat{\overline
{\partial}}=\delta/\delta\overline{\Theta}$\ are the functional derivatives
over the fields $\Theta$ and $\overline{\Theta}$, respectively. Thus, the
functional $H_{0}$ has the form of the energy%

\begin{equation}
H_{0}=\int d^{2}r\overline{\Theta}\left(  \frac{\overrightarrow{\widehat{p}%
}^{2}}{2m}-\mu+\frac{1}{2}g\overline{\Theta}\Theta\right)  \Theta\label{H0}%
\end{equation}

The integral over the fields $\theta$ and $\overline{\theta}$ in the
generating functional $Z$ (\ref{ZQCl0}) has the Gauss form and can easily be calculated%

\begin{equation}
Z=%
%TCIMACRO{\dint }%
%BeginExpansion
{\displaystyle\int}
%EndExpansion
D\Theta D\overline{\Theta}\exp\left(  -S\left[  \Theta\right]  +i\delta
S\left[  j\right]  \right)  \label{ZQCl}%
\end{equation}

The action $S\left[  \Theta\right]  $ can be written as%

\begin{equation}
S\left[  \Theta\right]  =\frac{1}{\Gamma_{K}}%
%TCIMACRO{\dint }%
%BeginExpansion
{\displaystyle\int}
%EndExpansion
dtd^{2}r\left\{  \left[  i\left(  1+i\varkappa\right)  \partial_{t}%
\Theta-F\right]  \left[  -i\left(  1-i\varkappa\right)  \partial_{t}%
\overline{\Theta}-\overline{F}\right]  \right\}  \label{SQCl}%
\end{equation}

Note that, in the case $\Gamma_{K}\rightarrow0$ due to the form of action
(\ref{SQCl}), the contribution to the functional integral (\ref{ZQCl}) is
given by the configurations of the field $\Theta$ obeying the standard
Gross-Pitaevskii equation. As is well known, the solution of this equation in
the assumption of the small magnitudes of momenta compared with the inverse
correlation length $\xi_{0}^{-1}\sim\sqrt{m\mu}$ yields the sound spectrum of
low energy excitations.

In the case $\Gamma_{K}\neq0$ the fluctuations of the Bose field $\Theta$
should be taken into account in the integral for the generating functional
(\ref{ZQCl}). Multiplying brackets in the action $S\left[  \Theta\right]  $
(\ref{SQCl}), we obtain%

\begin{equation}
S\left[  \Theta\right]  =\frac{1}{\Gamma_{K}}%
%TCIMACRO{\dint }%
%BeginExpansion
{\displaystyle\int}
%EndExpansion
dtd^{2}r\left\{
\begin{array}
[c]{c}%
\left(  1+\varkappa^{2}\right)  \left(  \partial_{t}\overline{\Theta}\right)
\left(  \partial_{t}\Theta\right)  +\overline{F}F+\\
+i\left(  \left(  \partial_{t}\overline{\Theta}\right)  F-\overline{F}\left(
\partial_{t}\Theta\right)  \right)
\end{array}
\right\}  \label{SmFr}%
\end{equation}

The complex fields $\Theta$ and $\overline{\Theta}$ can be rewritten in the
module-phase representation $\Theta=\rho\exp\left(  i\varphi\right)  $,
$\overline{\Theta}=\rho\exp\left(  -i\varphi\right)  $. In this representation
the corresponding terms of the action (\ref{SmFr}) take the form%

\begin{equation}
\left(  \partial_{t}\overline{\Theta}\right)  \left(  \partial_{t}%
\Theta\right)  =\rho^{2}\left(  \partial_{t}\varphi\right)  ^{2}+\left(
\partial_{t}\rho\right)  ^{2} \label{dTeta2}%
\end{equation}%
\begin{align}
\overline{F}F  &  =f^{2}\rho^{2}-\frac{1}{m}f\rho\left(  \overrightarrow
{\nabla}^{2}\rho-\rho\left(  \overrightarrow{\nabla}\varphi\right)
^{2}\right)  +\label{FF}\\
&  +\left(  \frac{1}{2m}\right)  ^{2}\left[
\begin{array}
[c]{c}%
\left(  \overrightarrow{\nabla}^{2}\rho\right)  ^{2}-2\rho\left(
\overrightarrow{\nabla}^{2}\rho\right)  \left(  \overrightarrow{\nabla}%
\varphi\right)  ^{2}+\\
+\rho^{2}\left(  \left(  \overrightarrow{\nabla}\varphi\right)  ^{2}\right)
^{2}+\rho^{2}\left(  \overrightarrow{\nabla}^{2}\varphi\right)  ^{2}+\\
+4\left(  \left(  \overrightarrow{\nabla}\rho\right)  \left(  \overrightarrow
{\nabla}\varphi\right)  \right)  ^{2}+4\rho\left(  \left(  \overrightarrow
{\nabla}\rho\right)  \left(  \overrightarrow{\nabla}\varphi\right)  \right)
\left(  \overrightarrow{\nabla}^{2}\varphi\right)
\end{array}
\right] \nonumber
\end{align}

where $f$ is%

\begin{equation}
f=g\rho^{2}-\mu\label{f}%
\end{equation}

and%

\begin{align}
&  i\left[  \left(  \partial_{t}\overline{\Theta}\right)  F-\overline
{F}\left(  \partial_{t}\Theta\right)  \right] \label{dTeta}\\
&  =2f\rho^{2}\partial_{t}\varphi+\frac{1}{m}\left(  \partial_{t}\rho\right)
\left[  2\left(  \overrightarrow{\nabla}\rho\right)  \left(  \overrightarrow
{\nabla}\varphi\right)  +\rho\overrightarrow{\nabla}^{2}\varphi\right]
-\nonumber\\
&  -\frac{1}{m}\left(  \rho\partial_{t}\varphi\right)  \left[  \overrightarrow
{\nabla}^{2}\rho-\rho\left(  \overrightarrow{\nabla}\varphi\right)
^{2}\right] \nonumber
\end{align}

The module $\rho$ of the field $\Theta$\ can be represented as a sum of the
average value and fluctuation of module $\rho=\rho_{0}+\delta\rho$, where
$\rho_{0}$ is the bare value of the average module \cite{Popov} $\rho
_{0}=\sqrt{\mu/g}$, and $\delta\rho$ is the module fluctuation. For small
frequencies $\omega<<\mu$ and momenta $\mid\overrightarrow{p}\mid<<\sqrt{m\mu
}$ the fluctuations $\delta\rho$ are small compared with the average
module$\ \delta\rho<<\rho_{0}$. The fluctuations of the phase in the 2D case
can be large \cite{Berez}, \cite{KT}, \cite{Popov}. For small module
fluctuations $\delta\rho<<\rho_{0}$,\ the function $f$ takes the form%

\begin{equation}
f=2g\rho_{0}\delta\rho=2\mu\delta\rho\label{df}%
\end{equation}

The expansion over the module fluctuations $\delta\rho$ for the small values
of the time and space derivatives $\left(  \frac{1}{2m}\overrightarrow{\nabla
}^{2}\delta\rho\right)  ,$ $\left(  \partial_{t}\delta\rho\right)  <<\mu
\rho_{0}$ leads to%

\begin{equation}
\overline{F}F=4\mu^{2}\delta\rho^{2}+2\mu\rho_{0}\delta\rho\left(  \frac{1}%
{m}\left(  \overrightarrow{\nabla}\varphi\right)  ^{2}\right)  +\rho_{0}%
^{2}\left(  \frac{1}{2m}\left(  \overrightarrow{\nabla}\varphi\right)
^{2}\right)  ^{2}+\rho_{0}^{2}\left(  \frac{1}{2m}\overrightarrow{\nabla}%
^{2}\varphi\right)  ^{2} \label{FF1}%
\end{equation}

\begin{equation}
i\left(  \left(  \partial_{t}\overline{\Theta}\right)  F-\overline{F}\left(
\partial_{t}\Theta\right)  \right)  =4\mu\rho_{0}\delta\rho\left(
\partial_{t}\varphi\right)  +2\rho_{0}^{2}\left(  \partial_{t}\varphi\right)
\left(  \frac{1}{2m_{pol}}\right)  \left(  \overrightarrow{\nabla}%
\varphi\right)  ^{2} \label{dTeta1}%
\end{equation}

\begin{equation}
\left(  \partial_{t}\overline{\Theta}\right)  \left(  \partial_{t}%
\Theta\right)  =\rho_{0}^{2}\left(  \partial_{t}\varphi\right)  ^{2}+\left(
\partial_{t}\delta\rho\right)  ^{2} \label{dTeta20}%
\end{equation}

The substitution of the above expressions into action (\ref{SmFr}) gives the
generating functional and the corresponding action%

\begin{equation}
Z=\int D\delta\rho D\varphi\exp\left\{  iS\left[  \delta\rho,\varphi\right]
+i\delta S_{\delta\rho,\varphi}\right\}  \label{ZdRodFai}%
\end{equation}

\begin{equation}
S\left[  \delta\rho,\varphi\right]  =\frac{1}{\Gamma_{K}}%
%TCIMACRO{\dint }%
%BeginExpansion
{\displaystyle\int}
%EndExpansion
dtd^{2}r\left\{
\begin{array}
[c]{c}%
\left(  1+\varkappa^{2}\right)  \rho_{0}^{2}\left(  \partial_{t}%
\varphi\right)  ^{2}+\\
+4\mu^{2}\delta\rho^{2}+4\mu\rho_{0}\delta\rho\left(  \frac{1}{2m}\left(
\overrightarrow{\nabla}\varphi\right)  ^{2}\right)  +\\
+\rho_{0}^{2}\left(  \frac{1}{2m}\left(  \overrightarrow{\nabla}%
\varphi\right)  ^{2}\right)  ^{2}+\rho_{0}^{2}\left(  \frac{1}{2m}%
\overrightarrow{\nabla}^{2}\varphi\right)  ^{2}+\\
+4\mu\rho_{0}\delta\rho\partial_{t}\varphi+2\left(  \rho_{0}^{2}\partial
_{t}\varphi\right)  \frac{1}{2m}\left(  \overrightarrow{\nabla}\varphi\right)
^{2}%
\end{array}
\right\}  \label{dS}%
\end{equation}

Integrating the generating functional $Z$ (\ref{ZdRodFai})\ over the module
fluctuations $\delta\rho$, we obtain the generating functional depending on
the phase field $\varphi$ alone%

\begin{equation}
Z=\int D\varphi\exp\left\{  iS\left[  \varphi\right]  +i\delta S_{\varphi
}\left[  j\right]  \right\}  \label{ZFi}%
\end{equation}

Here the effective action $S\left[  \varphi\right]  $ for the small values of
$\partial_{t}\varphi$ and $\frac{1}{2m}\overrightarrow{\nabla}^{2}\varphi
$\ compared with $\varkappa\mu$\ if $\varkappa<1$\ or $\mu$ if $\varkappa
>1$\ takes the form%

\begin{equation}
S\left[  \varphi\right]  =\frac{1}{\Gamma_{K}}%
%TCIMACRO{\dint }%
%BeginExpansion
{\displaystyle\int}
%EndExpansion
dtd^{2}r\left\{  \varkappa^{2}\rho_{0}^{2}\left(  \partial_{t}\varphi\right)
^{2}+\rho_{0}^{2}\left(  \frac{1}{2m}\overrightarrow{\nabla}^{2}%
\varphi\right)  ^{2}\right\}  \label{SFi}%
\end{equation}

and%

\[
\delta S_{\varphi}=\int dtd^{2}r\left[  \overline{j}\rho_{0}\exp\left(
i\varphi\right)  +j\rho_{0}\exp\left(  -i\varphi\right)  \right]
\]

The action (\ref{SFi}) describes a diffusion. It can be rewritten as%

\begin{equation}
S\left[  \varphi\right]  =\frac{\varkappa^{2}\rho_{0}^{2}}{\Gamma_{K}}%
%TCIMACRO{\dint }%
%BeginExpansion
{\displaystyle\int}
%EndExpansion
dtd^{2}r\left\{  \left(  \partial_{t}\varphi\right)  ^{2}+\left(
D_{0}\overrightarrow{\nabla}^{2}\varphi\right)  ^{2}\right\}  \label{SD0}%
\end{equation}

where $D_{0}$ is the diffusion coefficient%

\begin{equation}
D_{0}=\frac{1}{2m\varkappa} \label{D0}%
\end{equation}

The action (\ref{SD0}) gives the following expression for the phase correlator%

\begin{equation}
<\varphi_{\overrightarrow{k},\omega}\varphi_{-\overrightarrow{k}^{\prime
},-\omega^{\prime}}>=\frac{\left(  \Gamma_{K}/\varkappa^{2}\rho_{0}%
^{2}\right)  }{\omega^{2}+\left(  D_{0}\overrightarrow{k}^{2}\right)  ^{2}%
}\delta\left(  \overrightarrow{k}-\overrightarrow{k}^{\prime}\right)
\delta\left(  \omega-\omega^{\prime}\right)  \label{FiFi}%
\end{equation}

Near the pole this correlator can be rewritten as%

\begin{equation}
<\varphi_{\overrightarrow{k},\omega}\varphi_{-\overrightarrow{k}^{\prime
},-\omega^{\prime}}>=\frac{\left(  \Gamma_{K}/\varkappa^{2}\rho_{0}%
^{2}\right)  }{\left(  2iD_{0}\overrightarrow{k}^{2}\right)  \left(
\omega+iD_{0}\overrightarrow{k}^{2}\right)  }\delta\left(  \overrightarrow
{k}-\overrightarrow{k}^{\prime}\right)  \delta\left(  \omega-\omega^{\prime
}\right)  \label{FiFi1}%
\end{equation}

The correlator describes the properties of low energy excitations. For the
thermodynamically equilibrium Bose system, the phase correlator has the real
pole of the sound type for small momenta and frequencies. In contrast for
$\Gamma_{K}\neq0$, the phase correlator (\ref{FiFi1}) has the diffusion pole
at the small momenta and frequencies. This form of the phase correlator means
the absence of the phase coherence and, as a result, the lack of BEC and
superfluidity in the stationary state of nonequilibrium Bose gas with the
dynamic balance between the particle escape and creation processes. For
example, in the case of the stationary state of nonequilibrium Bose gas the
quantum interference experiments, like those in trapped quantum atomic Bose
gases on the interference of two independent Bose condensates \cite{AnKett},
\cite{Pit}, cannot give the pronounced interference picture. Note that the
derivation of the low energy phase correlator (\ref{FiFi}) does not use the
D=2 dimensionality of the system and can be used without any change for
nonequilibrium D=3 Bose gas in the stationary state.

The phase correlator at the equal time moments can easily be obtained from Eq.
(\ref{FiFi})
\begin{equation}
<\varphi_{\overrightarrow{k},t}\varphi_{-\overrightarrow{k},t}>=\left(
\Gamma_{K}/\varkappa^{2}\rho_{0}^{2}\right)  \frac{1}{2D_{0}\overrightarrow
{k}^{2}}=\frac{\Gamma_{K}m}{\varkappa\rho_{0}^{2}}\frac{1}{\overrightarrow
{k}^{2}} \label{FiFit}%
\end{equation}

The substitution of Eq. (\ref{Kap}) into Eq. (\ref{FiFit}) gives the phase
correlator in the form%

\begin{equation}
<\varphi_{\overrightarrow{k},t}\varphi_{-\overrightarrow{k},t}>=\frac
{4T^{\ast}m}{\rho_{0}^{2}}\frac{1}{\overrightarrow{k}^{2}} \label{FiFi2}%
\end{equation}

This form of the phase correlator is typical for the interacting Bose gas at
low temperature. However, it results from the diffusion behavior of the
excitation spectrum and not from the sound behavior of the spectrum.

The transition to the space coordinates in the 2D case leads to the typical
expression for the Bose field correlator at large distances and time moments
\cite{Berez}, \cite{KT}%

\begin{equation}
<\overline{\Theta}_{\overrightarrow{r},t}\Theta_{\overrightarrow{r}^{\prime
},t^{\prime}}>=\rho_{0}^{2}\exp\left(  -\frac{4T^{\ast}m}{\rho_{0}^{2}}%
\ln\left(  \frac{s}{\xi_{0}}\right)  \right)  \label{BKT}%
\end{equation}

where $\xi_{0}^{-1}=\sqrt{mg\rho_{0}^{2}}$, $s=\sqrt{\mid\overrightarrow
{r}-\overrightarrow{r}^{\prime}\mid^{2}+D_{0}\mid\left(  t-t^{\prime}\right)
\mid}$ and $\xi_{0}$ is the effective correlation length.

It should be noted that the behavior of the system in the stationary state
differs from the behavior of the equilibrium Bose gas in the
Berezinskii-Kosterlitz-Thouless state. In the equilibrium Bose gas there is a
local condensate with the coherent phase in the regions which size is much
smaller than the average distance between vortices and larger than $\xi_{0}$.
The local condensate with the coherent phase in the stationary state of the
nonequilibrium Bose system is absent. In this case the phase correlator has a
diffusion character on all scales larger than $\xi_{0}$.

\section{Conclusion.}

In the present paper the low energy properties of the stationary state of
nonequilibrium Bose gas have been considered at sufficiently low temperatures
$T^{\ast}<<\mu$ . In the stationary state the particle escape and creation
compensate each other in average, i.e., the stationary state requires zero
value of relaxations $\Gamma_{R}\left(  \omega\right)  $ and $\Gamma
_{A}\left(  \omega\right)  $ for zero frequency $\omega=0$, where
$\Gamma_{R,A}\left(  \omega\right)  =\Gamma_{R,A}^{\left(  out\right)
}\left(  \omega\right)  -\Gamma_{R,A}^{\left(  in\right)  }\left(
\omega\right)  $. However, the quantum noise of the Bose field at zero
frequency does not vanish and is characterized by $\Gamma_{K}=\Gamma
_{K}^{\left(  out\right)  }\left(  0\right)  +\Gamma_{K}^{\left(  in\right)
}\left(  0\right)  $, where the kinetic component of the self-energy part
$\Gamma_{K}$ at zero frequency is a sum of positive values $\Gamma
_{K}^{\left(  out\right)  }\left(  0\right)  $ and $\Gamma_{K}^{\left(
in\right)  }\left(  0\right)  $. Due to this noise the low energy spectrum of
elementary excitations, which is governed by the phase fluctuations of the
Bose field, has the diffusion character (\ref{FiFi1}). The diffusion character
of the phase fluctuations means the absence of the phase coherency of the
classical Bose field and, therefore, the absence of BEC in the stationary
state of nonequilibrium Bose system. The diffusion character of the low energy
spectrum of elementary excitations results from nonzero value of the zero
frequency component of $\Gamma_{K}\left(  \omega\right)  $. Note that the
behavior of the module of the Bose field in the stationary nonequilibrium
state at small effective temperatures is the same as in the thermodynamically
equilibrium state, i.e., module of the Bose field has homogeneous nonzero
average value and the field fluctuations are small compared with the average
value. Moreover, the momentum behavior of the correlator of the phase
fluctuations in the nonequilibrium stationary state is analogous to that in
the thermodynamically equilibrium state with the Bose condensate.

Acknowledgment. The author is grateful to V.M. Agranovich for attracting
author's attention to the problem and for useful discussions.

\end{document}